\documentstyle[preprint,eqsecnum,aps]{revtex}
\tightenlines                                                                                                                                                                
\newcommand{\be}{\begin{equation}}
\newcommand{\ee}{\end{equation}}
\newcommand{\bea}{\begin{eqnarray}}
\newcommand{\eea}{\end{eqnarray}}
\newcommand{\ba}{\begin{array}}
\newcommand{\ea}{\end{array}}

\newcommand{\th}{\theta}
\newcommand{\Ga}{\Gamma}

\newcommand{\la}{\lambda}
\newcommand{\La}{\Lambda}

\newcommand{\pa}{\partial}
\newcommand{\pax}{\partial_x}

\newcommand{\no}{\nonumber}

\newcommand{\Om}{\Omega}

\newcommand{\bchi}{\mbox{\boldmath${\chi}$}}

\newcommand{\sdet}{\mbox{sdet}}
\newcommand{\sres}{\mbox{sres}}

\newcommand{\br}{\breve}
\newcommand{\ti}{\tilde}
\newcommand{\lb}{\label}

\begin{document}

\title{Binary Darboux-B\"acklund Transformations for \\
the Manin-Radul Super KdV Hierarchy}

\author{Jiin-Chang Shaw$^1$ and Ming-Hsien Tu$^2$ }
\address{
$^1$ Department of Applied Mathematics, National Chiao Tung University, \\
Hsinchu, Taiwan, Republic of China,\\
and\\
$^2$ Department of Physics, National Tsing Hua University, \\
Hsinchu, Taiwan, Republic of China
}
\date{\today}
\maketitle

\begin{abstract}
\end{abstract}
We construct the supersymmetric extensions of the Darboux-B\"acklund 
transformations (DBTs) for the Manin-Radul super KdV hierarchy using 
the super-pseudo-differential operators. The elementary DBTs are triggered 
by the gauge operators constructed from the wave functions and adjoint 
wave functions of the hierarchy. Iterating these elementary DBTs, we obtain 
not only Wronskian type but also binary type superdeterminant representations 
of the solutions. 
\newpage

\section{Introduction}

In the past few decades, the Darboux-B\"acklund transformations (DBTs) have 
been shown to be an efficient method to obtain the soliton solutions of the
nonlinear integrable systems including 
Korteweg-de Vries (KdV), Kadomtsev-Petviashili (KP), sine-Gordon and nonlinear
Schr\"odinger equations (see, for example Ref.\cite{MS}) etc. In contrast to
these results, the application of the DBTs to super integrable systems is quite limited. 
The reasons are partly due to the complexity of the supersymmetric formulation of 
super integrable systems. However, as far as the super soliton solutions are
concerned, the supersymmetric generalization of the DBTs is urgently needed.

Recently, Liu \cite{L} proposed a DBT for the Manin-Radul super KdV (MR sKdV)
equation \cite{MR}.
The DBT is triggered by a gauge operator which is a first order super-differential
operator (SDO) parametrized by a wave function of the associated linear system. By
iterating such DBT, Liu and Ma\~nas \cite{LM1} obtained the 
supersymmetric version of the Crum transformation for the MR sKdV
equation and showed that the solutions can be expressed in terms of 
Wronskian superdeterminants. 

In this paper, we will investigate the DBTs for the MR sKdV hierarchy
from super-pseudo-differential operator (S$\Psi$DO) point of view.
Motivated by the non-supersymmetric cases 
\cite{MS,CSY,OS,NI,ANP1,ANP,CST,LW,ST} 
and the works described above,  we introduce the adjoint DBT which is 
triggered by an adjoint wave function of the system. 
This adjoint DBT in combination with the previous one form a binary DBT 
which is constructed from a wave function and an adjoint wave function.
Iterating these elementary DBTs, we obtain not only Wronskian type but also
binary type superdeterminant representations for the solutions of the MR sKdV
hierarchy which have not been obtained previously. 

Our paper is organized as follows: In Sec. II, some basic facts about S$\Psi$DO
are recalled and the MR sKdV hierarchy is defined. In Sec. III, we introduce the
elementary DBTs which are triggered by the gauge operators
constructed from the wave functions and the adjoint wave functions of the
MR sKdV system. In Sec. IV, we iterate these DBTs to obtain
the solutions of the MR sKdV hierarchy. Concluding remarks are
presented in Sec. V.

\section{Preliminaries}
Before considering a specific super integrable system, let us recall
some basic facts of the S$\Psi$DO
which is defined by
\be
\La=D^{N}+\sum^{N-1}_{i=-\infty}U_iD^i \qquad N\in {\bf Z\/}
\ee
where the supercovariant derivative $D\equiv\pa_{\th}+\th\pa$ satisfies
$D^2=\pa$, $\th$ is the Grassmann variable ($\th^2=0$) which, together with
the even variable $x\equiv t_1$, defines 
the $(1|1)$ superspace with coordinate $(x,\th)$.
The formal inverse of $D$ is introduced by $D^{-1}=\th+\pa_{\th}\pa^{-1}$
which satisfies $DD^{-1}=D^{-1}D=1$.
The coefficients $U_i$ are superfields that depend on the 
variables $x$, $\th$, and $t_i$ and can be represented by $U_i=u_i(t)+\th v_i(t)$. 
Since the S$\Psi$DO is assumed to be
homogeneous under $Z_2$-grading, the grading of the superfield $U_i$ is
$|U_i|=N+i$ (mod 2). Here we refer the parity of $U_i$ to be even if $|U_i|=0$
and odd if $|U_i|=1$.
The supercovariant derivative $D$ satisfies the 
supersymmetric version of the Leibniz rule \cite{MR}:
\be
D^iU=\sum_{k=0}^{\infty}(-1)^{|U|(i-k)} {i\brack k}D^{k}(U)D^{i-k}
\label{leib}
\ee
where the super-binomial coefficients ${i\brack k}$ are defined by
\be
{i\brack k}=
\left\{
\ba{l}
{[i/2]\choose [k/2]} \qquad \mbox{for $ 0\leq k\leq i$ and $(i,k)\ne (0,1)$ mod 2}\\
(-1)^{[k/2]}{-i+k-1\brack k}\qquad \mbox{for $i<0$}\\
0\qquad \mbox{otherwise}
\ea
\right.
\ee
It is convenient to separate $\La$ into the positive and negative parts as follows:
\be
\La_+=\sum_{i\geq 0}U_iD^i, \qquad \La_-=\sum_{i\leq -1}U_iD^i.
\ee
The super-residue ($\sres$) of $\La$ is defined by
\be
\sres(\La)=U_{-1}
\ee
and the conjugate operation ``$\ast$" for 
S$\Psi$DOs can be defined as follows
\be
(PQ)^*=(-1)^{|P||Q|}Q^*P^*
\ee
which implies that $U^*=U$ for arbitrary superfield $U$ and
\be
(\pa^i)^*=(-1)^i\pa^i,\qquad (D^i)^*=(-1)^{i(i+1)/2}D^i.
\ee
A simple calculation shows that
\be
\La^*=\sum_i(-1)^N(-1)^{i(i-1)/2}D^iU_i
\ee
With these definitions in hand, let us provide some useful 
identities which will simplify the computations  involving 
compositions of S$\Psi$DOs.

{\bf Lemma 1 :\/}
\bea
& &(\La^*)_+=(\La_+)^*\qquad (\La^*)_-=(\La_-)^* \\
& &(D^{-1}\La)_-=D^{-1}(\La^*)_0+D^{-1}(\La)_-\\
& &(\La D^{-1})_-=(\La)_0D^{-1}+\La_-D^{-1}\\
& &\sres(\La)=\sres(\La^*)\qquad (D\sres(\La))=\sres(D\La-(-1)^{|\La|}\La D)\\
& &\sres(\La D^{-1})=(\La)_0\qquad \sres(D^{-1}\La)=(-1)^{|\La|}(\La^*)_0\\
& &\sres(D^{-1}\La_1\La_2D^{-1})
=\sres(D^{-1}(\La_1^*)_0\La_2D^{-1})+\sres(D^{-1}\La_1(\La_2)_0D^{-1})
\label{id4}
\eea
where $\La_1=(\La_1)_+$ and $\La_2=(\La_2)_+$.

{\it Proof.\/} The proofs for these identities are straightforward. Here we only give the
proof for the second identity. From the left-hand side, we have
\be
(D^{-1}\La)_-=(D^{-1}\La_+)_-+D^{-1}\La_-.
\ee
Then using (\ref{leib}), we obtain 
\bea
(D^{-1}\La_+)_-&=&(\sum_{k=0}^{\infty}\sum_{i=0}^{N}(-1)^{(N+i)(1+k)}
{-1 \brack k}(D^kU_i)D^{-1-k+i})_-\no\\
&=&\sum_{k=0}^{\infty}\sum_{i=0}^{N}(-1)^{(N+i)(1+i+l)}
(-1)^{[(i+l)/2]}(D^{i+l}U_i)D^{-1-l}\no\\
&=&\sum_{l=0}^{\infty}(-1)^{N(1+l)}(-1)^{[l/2]}(D^l(\La^*)_0)D^{-1-l}\no\\
&=&D^{-1}(\La^*)_0
\label{pf1}
\eea
here the relation $(-1)^{[i/2]}=(-1)^{i(i-1)/2}$ has been used to reach the third
line of (\ref{pf1}).$ \Box$

The MR sKdV hierarchy is defined, 
in Lax form, as
\be
\pa_{t_n} L=[P_n,L]\qquad n=1,3,5\cdots
\lb{leq}
\ee 
with
\bea
L&=&\pa^2+v_1D+v_0
\lb{l}\\
P_n&=&L^{n/2}_+
\lb{l3}
\eea
where the coefficients $v_1$ and $v_0$ are superfields depending on the variables
$(\th,x,t_3,t_5,\cdots)$ with grading $|v_i|=i$ (mod 2).
We can rewrite the hierarchy equations (\ref{leq}) as follows :
\be
\pa_{t_m}P_n-\pa_{t_n}P_m+[P_n,P_m]=0
\label{zc}
\ee
which is called the zero-curvature condition and is equivalent to the whole
set of equations of (\ref{leq}). If we can find a set of superfields $\{v_1,v_0\}$
and hence a corresponding set of super-differential operator (SDO) $\{P_n\}$
satisfying (\ref{zc}), then we have a solution to the MR sKdV hierarchy.
In fact, the MR sKdV hierarchy was constructed originally from the 
MR sKP hierarchy \cite{MR} associated with the odd S$\Psi$DO: 
$\La_{MR}=D+\sum_{i=0}^{\infty}U_{-i}D^{-i}$
by the reduction $L=(\La_{MR}^4)_+$ and has been shown to 
be bi-Hamiltonian\cite{OP,FR}.
Substituting (\ref{l}) and (\ref{l3}) into the Lax
equation (\ref{leq}) for $n=3$, one obtains
\bea
\pa_{t_3}v_0&=&\frac{1}{4}\pax(v_{0xx}+3v_0^2+3v_1(Dv_0)),\qquad 
\lb{eq1}\\
\pa_{t_3}v_1&=&\frac{1}{4}\pax(v_{1xx}+3v_1(Dv_1)+6v_1v_0)
\lb{eq2}
\eea
which is the MR sKdV equation.
By setting $v_1=0$, Eqs. (\ref{eq1})-(\ref{eq2}) reduce to the KdV equation. 
The Lax equation (\ref{leq}) can be viewed as
the compatibility condition of the linear system
\be
L\phi=\la\phi, \qquad \pa_{t_n}\phi=(P_n\phi)_0
\lb{wf}
\ee
where $\phi$ and $\la$ are called wave function and spectral parameter
of the hierarchy, respectively. On the other hand, we can also introduce
adjoint wave function $\psi$ which satisfies the linear system
\be
L^*\psi=\eta\psi, \qquad \pa_{t_n}\psi=-((P_n)^*\psi)_0
\lb{awf}
\ee
For convenience, throughout this paper, $\phi$ and $\psi$ will
stand for wave function and adjoint wave function, respectively.
Of course, it should be realized that both $\phi$ and $\psi$
are superfields in this formalism.
In next section, we will use these (adjoint) wave functions to
construct the DBTs for the MR sKdV hierarchy.

\section{The elementary DB transformations}

We consider the following transformation:
\be
L\rightarrow \hat{L}=TLT^{-1}
\lb{tran}
\ee
where $T=T(\th,x,t_3,t_5,\cdots)$ is any reasonable S$\Psi$DO. 
To guarantee that such transformation can generate new
solutions of the MR sKdV hierarchy, the transformed Lax operator $\hat{L}$ 
should preserve the form of (\ref{l}) and satisfy the Lax equation (\ref{leq}).  
It is easy to show that under the transformation (\ref{tran}), 
the operator $P_n$ then is transformed as
\be
P_n\rightarrow \hat{P}_n=TP_nT^{-1}+\pa_{t_n}TT^{-1}.
\lb{p3}
\ee
which preserves the zero curvature condition (\ref{zc}).
Note that although $P_n$ is a SDO, the right hand side
of (\ref{p3}) will in general not be a purely SDO.
However if we suitable choose the gauge operator $T$ such that
$ \hat{P}_n$, as defined by (\ref{p3}), is also a purely SDO,
then $\{\hat{L},\hat{P}_n\}$ represents a valid new solution to the MR sKdV
hierarchy. To formulate the DBTs  of the MR sKdV hierarchy,
let us introduce a superfield $\Om$ called bi-linear potential \cite{OS}
(or squared eigenfunction potential) which is constructed from a wave function and
an adjoint wave function  and will be useful later on.

{\bf Lemma 2 :\/} For any pair of $\phi$ and $\psi$, there exists a
bi-linear potential $\Om(\psi,\phi)$ satisfying
\bea
(D\Om(\psi,\phi))&=&\psi\phi,
\lb{bi1}\\
\pa_{t_n}\Om(\psi,\phi)&=&\sres(D^{-1}\psi P_n\phi D^{-1}),
\lb{bi2}
\eea

{\it Proof.\/} To prove the existence of the bi-linear potential $\Om$, we 
have to show that Eqs.(\ref{bi1}) and (\ref{bi2}) are compatible. 
Firstly, we note that (\ref{bi1}) is consistent with (\ref{bi2}) for $n=1$. Since
\bea
\Om_x&=&\sres(D^{-1}\psi P_1\phi D^{-1}) \no\\
&=&\sres(D^{-1}\psi D^2\phi D^{-1})\no\\
&=&(D\psi)\phi+(-1)^{|\psi|}\psi(D\phi)\no\\
&=&(D(D\Om)).
\eea
This implies that
\be
(D\Om)_x=(D(D(D\Om)))=(D\Om_x).
\ee
In general, from (\ref{bi1}) and (\ref{bi2}) we have
\bea
(D\pa_{t_n}\Om)&=&(D\sres(D^{-1}\psi P_n\phi D^{-1}))\no\\
&=&\sres(\psi P_n\phi D^{-1}-(-1)^{|\psi||\phi|}D^{-1}\psi P_n\phi)\no\\
&=&\psi(P_n\phi)-(-1)^{|\psi||\phi|}\phi(P_n^*\psi)\no\\
&=&\psi\phi_{t_n}+\psi_{t_n}\phi\no\\
&=&\pa_{t_n}(D\Om).
\eea
Secondly, from (\ref{bi2}) and (\ref{id4}) we have
\bea
\pa_{t_m}(\pa_{t_n}\Om)-\pa_{t_n}(\pa_{t_m}\Om)
&=&\sres\pa_{t_m}(D^{-1}\psi P_n\phi D^{-1})-
\sres\pa_{t_n}(D^{-1}\psi P_m\phi D^{-1})\no\\
&=&\sres(D^{-1}\psi (\pa_{t_m}P_{n}-\pa_{t_n}P_{m})\phi D^{-1})\no\\
& &-\sres(D^{-1}\psi (P_m^*\psi)P_n\phi D^{-1})+
\sres(D^{-1}\psi P_n(P_m\phi)D^{-1})\no\\
& &+\sres(D^{-1}\psi (P_n^*\psi)P_m\phi D^{-1})-
\sres(D^{-1}\psi P_m(P_n\phi)D^{-1})\no\\
&=&\sres(D^{-1}\psi(\pa_{t_m}P_{n}-\pa_{t_n}P_{m}+
[P_n,P_m])\phi D^{-1})=0
\eea
here the zero-curvature condition of the MR sKdV hierarchy has been used in the
last line.$\Box$

{\bf Remarks.\/} $\Om(\psi,\phi)$ is a superfield with parity $(-1)^{|\psi||\phi|+1}$ 
and can be represented by 
$\Om(\psi,\phi)=\int^x\psi_2\phi_1+(-1)^{|\psi|}\int^x\psi_1\phi_2+\th\psi_1\phi_1$
where $\phi=\phi_1+\th\phi_2$ and $\psi=\psi_1+\th\psi_2$.

{\bf Proposition 1 :\/}\cite{L} Let $\chi$ be an even wave function of the linear system, 
then the gauge operator
\be
T=\chi D \chi^{-1}=D+\alpha \qquad \alpha\equiv -\frac{(D\chi)}{\chi}
\ee
triggers the following DBT:
\bea
\hat{L}&=&TLT^{-1}=\pa^2+\hat{v}_1D+\hat{v}_0,
\lb{t1}\\
\hat{\phi}&=&(T\phi)=\chi(D\chi^{-1}\phi),
\lb{t2}\\
\hat{\psi}&=&((T^{-1})^*\psi)=\chi^{-1}\Om(\psi,\chi),
\lb{t3}
\eea 
where the transformed coefficients are given by
\bea
\hat{v}_1&=&-v_1-2\alpha_x,
\lb{tv1}\\
\hat{v}_0&=&v_0+(Dv_1)+2\alpha(v_1+\alpha_x).
\lb{tv0}
\eea
{\it Proof.\/} We first show that $\hat{L}$ has the same form as $L$. 
From (\ref{t1}) we have
\be
\hat{L}_-=(\chi D\chi^{-1}L\chi D^{-1}\chi^{-1})_-
=\chi(D\chi^{-1}(L\chi))D^{-1}\chi^{-1}
=0.
\ee
which implies that $\hat{L}$ is a SDO. A simple calculation
shows that the transformed coefficients $\hat{v}_1$ and $\hat{v}_0$ 
are given by (\ref{tv1}) and (\ref{tv0}), respectively.
Furthermore, the hierarchy equation for $\hat{L}$ is given by
\be
\pa_{t_n}\hat{L}=[TP_nT^{-1}+\pa_{t_n}TT^{-1},\hat{L}].
\ee
Using the identity
\be
\chi D\chi^{-1}\La_+\chi D^{-1}\chi^{-1}=
(\chi D\chi^{-1}\La\chi D^{-1}\chi^{-1})_++
\chi (D\chi^{-1}(\La_+\chi)_0)D^{-1}\chi^{-1}
\label{identity}
\ee 
and then substituting $\La=L^{n/2}$ ($n$ odd), we obtain
\be
\hat{P}_n
=TP_nT^{-1}+\pa_{t_n}TT^{-1}
=(TL^{n/2}T^{-1})_++\chi(D\chi^{-1}\pa_{t_n}\chi)D^{-1}\chi^{-1}+
\pa_{t_n}TT^{-1}=\hat{L}^{n/2}_+
\ee
Hence
\be
\pa_{t_n}\hat{L}=[\hat{L}^{n/2}_+,\hat{L}]
\label{newleq}
\ee
The evolution equations for $\hat{\phi}$ and $\hat{\psi}$ can be verified in a 
similar way. From (\ref{t2}), we have
\be
\pa_{t_n}\hat{\phi}=\pa_{t_n}T\phi+T\pa_{t_n}\phi=
\pa_{t_n}TT^{-1}\hat{\phi}+TP_nT^{-1}\hat{\phi}=(\hat{P}_n\hat{\phi})_0.
\ee
Similarly,
\be
\pa_{t_n}\hat{\psi}=\pa_{t_n}(T^{-1})^*\psi+(T^{-1})^*\pa_{t_n}\psi
=((T^{-1})^*(\pa_{t_n}T)^*+(T^{-1})^*P_n^*T^*)\hat{\psi}=-
(\hat{P}^*_n\hat{\psi})_0
\qquad\Box
\ee
Having described the first construction of the DBT, we now turn to 
another construction using adjoint wave function.

{\bf Proposition 2 :\/} Let $\mu$ be an even adjoint wave function of the linear system 
(\ref{awf}), then the gauge operator
\be
S=\mu^{-1} D^{-1} \mu=(D-\beta)^{-1} \qquad \beta\equiv -\frac{(D\mu)}{\mu}
\ee
triggers the following adjoint DB transformation:
\bea
\hat{L}&=&SLS^{-1}=\pa^2+\hat{v}_1D+\hat{v}_0,\\
\hat{\phi}&=&(S\phi)=\mu^{-1}\Om(\mu,\phi),\\
\hat{\psi}&=&((S^{-1})^*\psi)=-\mu(D\mu^{-1}\psi),
\eea 
where
\bea
\hat{v}_1&=&-v_1+2\beta_x,\\
\hat{v}_0&=&v_0+(Dv_1)-2(D\beta)_x-2\beta(v_1-\beta_x).
\eea
{\it Proof.\/} The proof is similar to the Proposition 1. Here we only mention that
Eq.(\ref{newleq}) can be verified easily by using the identity
\be
(\mu^{-1}D^{-1}\mu\La\mu^{-1}D\mu)_-=
\mu^{-1}D^{-1}\mu\La_-\mu^{-1}D\mu-(-1)^{|\La|}
\mu^{-1}D^{-1}(D\mu^{-1}(\La_+^*\mu)_0)\mu
\ee
instead of (\ref{identity}).
$\Box$

{\bf Proposition 3 :\/} For any pair of even wave function $\chi$ and odd adjoint 
wave function $\mu$, the gauge operator
\be
R=1-\chi\Om(\mu,\chi)^{-1}D^{-1}\mu
\lb{r}
\ee
triggers the following binary DB transformation:
\bea
\hat{L}&=&RLR^{-1}=\pa^2+\hat{v}_1D+\hat{v}_0,\\
\hat{\phi}&=&(R\phi)=\phi-\chi\Om(\mu,\chi)^{-1}\Om(\mu,\phi),\\
\hat{\psi}&=&((R^{-1})^*\psi)=\psi-\mu\Om(\mu,\chi)^{-1}\Om(\psi,\chi),
\eea 
where
\bea
\hat{v}_1&=&v_1-2\Ga_x \qquad \Ga\equiv \frac{(D\Om)}{\Om}
\lb{rv1}\\
\hat{v}_0&=&v_0+2(D\Ga)_x+2v_1\Ga-2\Ga \alpha_x+2\alpha\Ga_x+2\Ga\Ga_x.
\lb{rv0}
\eea
{\it Proof.\/} The gauge operator $R$ is just a composition of the DB transformation
$T$ and the adjoint DB transformation $S$. To see this, let us first perform a DB 
transformation
triggered by the wave function $\chi$. The odd adjoint wave function $\mu$ is thus 
transformed to $\hat{\mu}=\chi^{-1}\Om(\mu,\chi)$ which is an even one. 
Then a subsequent adjoint DB
transformation $S$ triggered by $\hat{\mu}$ is performed and the composition
of these two transformations is given by $
R=\hat{\mu}^{-1}D^{-1}\hat{\mu}\chi D\chi^{-1} 
=\chi\Om^{-1}D^{-1}\Om D\chi^{-1}
=1-\chi\Om^{-1}D^{-1}\mu$.
Therefore, the remaining part of the proof is just a corollary of the 
Propositions 1 and 2.$\Box$

{\bf Remarks.\/} The binary DB transformation can also be constructed from a
pair of $\chi$ and $\mu$ with $|\chi|=1$ and $|\mu|=0$. A direct calculation
shows that 
$R=\hat{\chi}D\hat{\chi}^{-1}\mu^{-1} D^{-1}\mu= 1-\chi\Om^{-1}D^{-1}\mu$
which has the same form as (\ref{r}).
We also note that the transformed coefficients (\ref{rv1}) and (\ref{rv0}) 
are just the ones in Ref.\cite{L} where a different approach was presented.

\section{Iteration of the DB transformations}

We have introduced the elementary DBTs which contain
the DBT, the adjoint DBT, and the binary 
DBT triggered by the gauge operators $T$, $S$, and $R$, respectively. 
Using these elementary transformations as the building blocks, 
the more complicated transformations can be
constructed from the compositions of these gauge operators. 
In Ref.\cite{LM1}, by iterating the DB transformation $T$, the so-called 
Crum transformation for  the MR sKdV equation was constructed 
and  the Wronskian superdeterminant representations for the solutions were obtained. 
This construction starts with $n$ wave functions $\chi_i, i=0,\cdots,n-1$ of
the linear system (\ref{wf}) with parity $(-1)^i$. 
Using $\chi_0$, say, to perform the first DBT of Proposition 1, 
then $\chi_i$ are transformed to $\hat{\chi}_i$.  It is obvious that $\hat{\chi}_0=0$.
The next step is to perform a subsequent DBT triggered by
$\hat{\chi}_1$, say,  which leads to the new wave functions $\hat{\hat{\chi}}_i$
with $\hat{\hat{\chi}}_1=0$. Iterating this process such that 
all the wave functions are used up, then an $n$-step DBT with
gauge operator $T_n$ is obtained.

{\bf Proposition 4 :\/}\cite{LM1} Let $\chi_i, i-0,\cdots,n-1$ be wave functions 
of the linear system (\ref{wf}) with parity $(-1)^i$, then after $n$ iterations 
of the DBT of Proposition 1, the transformed Lax operator becomes
\bea
\hat{L}&=&T_nLT_n^{-1} \qquad T_n=D^n+\sum_{i=0}^{n-1}a_iD^i\\
&=&\pa^2+\hat{v}_1D+\hat{v}_0
\eea
where the coefficients $a_i$ are defined by
\be
T_n\chi_j=0\qquad j=0,\cdots,n-1
\lb{lintn}
\ee
and
\bea
\hat{v}_1&=&(-1)^nv_1-2(a_{n-1})_x\\
\hat{v}_0&=&v_0-2(a_{n-2})_x-a_{n-1}((-1)^nv_1+\hat{v}_1)+
\frac{1-(-1)^n}{2}(Dv_1)
\eea
{\bf Remarks.\/} Due to the fact that the parity of the gauge operator $T_n$
depends on $n$, the cases for $n$ being even or odd are rather different. In Ref.\cite{LM1},
the coefficients $a_i$ for both cases have been obtained separately by solving
the linear equation (\ref{lintn}) and the transformed fields $\hat{v}_1$ and $\hat{v}_0$
can be expressed in a compact form in terms of superdeterminant. We  will
present such calculation in a more general case involving the binary
DBTs. 

Now let us turn to a similar construction using
the adjoint wave functions.

{\bf Proposition 5 :\/} Let $\mu_i, i=0,\cdots,n-1$ be adjoint wave functions of 
the linear system (\ref{awf}) with parity $(-1)^i$, then after $n$ iterations of the 
adjoint DBT of Proposition 2, the transformed Lax operator becomes
\bea
\hat{L}&=&S_nLS_n^{-1} \qquad 
(S_n^{-1})^*=(-1)^{[n/2]}(D^n+\sum_{i=0}^{n-1}b_iD^i)\\
&=&\pa^2+\hat{v}_1D+\hat{v}_0
\eea
where the coefficients $b_i$ are defined by
\be
(S_n^{-1})^*\mu_j=0\qquad j=0,\cdots,n-1
\ee
and
\bea
\hat{v}_1&=&(-1)^nv_1+2(b_{n-1})_x\\
\hat{v}_0&=&v_0-2(b_{n-2})_x+b_{n-1}((-1)^nv_1+\hat{v}_1)+
\frac{1+(-1)^n}{2}(Dv_1)-(D\hat{v}_1)
\eea
{\it Proof.\/} Essentially, the proof is exactly the same as Proposition 4.$\Box$

{\bf Proposition 6 :\/} Let $\chi_0,\cdots,\chi_{n-1}$ be even wave functions of the
linear system (\ref{wf}) and $\mu_0,\cdots,\mu_{n-1}$ be odd 
adjoint wave functions of the linear system (\ref{awf}). 
Then after $n$ iterations of the binary DBT of Proposition 3, 
the transformed Lax operator becomes
\bea
\hat{L}&=&R_nLR_n^{-1} \qquad R_n=1-\sum_{i=0}^{n-1}c_iD^{-1}\mu_i\\
&=&\pa^2+\hat{v}_1D+\hat{v}_0
\eea
where the coefficient $c_i$ are defined by
\be
R_n\chi_j=0 \qquad j=0,\cdots,n-1
\ee 
and 
\bea
\hat{v}_1&=&v_1-2\sum_{i=0}^{n-1}(c_i\mu_i)_x
\lb{rnv1}\\
\hat{v}_0&=&v_0+(v_1+\hat{v}_1)\sum_{i=0}^{n-1}c_i\mu_i
+2\sum_{i=0}^{n-1}(c_i(D\mu_i))_x
\lb{rnv0}
\eea
{\it Proof.\/} A straightforward calculation. We only remark that
the form of the gauge operator $R_n$ can be verified by induction. $\Box$

{\bf Proposition 7 :\/} The transformed coefficients, $\hat{v}_1$ and $\hat{v}_0$, 
in Eqs.(\ref{rnv1})-(\ref{rnv0}) can be expressed as
\bea
\hat{v}_1&=&v_1-2(D^3\ln\det\Om)\\
\hat{v}_0&=&v_0+(v_1+\hat{v}_1)(D\ln\det\Om)+
2\sum_{i=0}^{n-1}(D^2(\frac{\det\Om^{(i)}}{\det\Om}(D\mu_i)))
\eea
where $\Om_{ij}\equiv \Om(\mu_i,\chi_j)$ and $\Om^{(i)}$ is constructed from
$\Om$ with its $i$-th row replaced by 
($\chi_0,\cdots,\chi_{n-1}$). 

{\it Proof.\/} The proof follows easily from the Cramer's formula. $\Box$

{\bf Remarks.\/} The expressions of the transformed fields $\hat{v}_1$ 
and $\hat{v}_0$ show that they are unchanged under the interchange 
of any two wave functions $\chi_i$ and $\chi_j$ with $i\ne j$. Hence
the permutability of DBTs is still maintained in this 
supersymmetric formalism. We also note that the case for 
$|\chi_i|=1$ and $|\mu_i|=0$ gives the same result.

Finally let us discuss a more general DBT in which
the numbers of wave functions and adjoint wave functions are unequal.

{\bf Proposition 8 :\/} Suppose there are $n$ adjoint wave functions 
$\mu_0,\cdots,\mu_{n-1}$ and $n+m$
wave functions $\chi_0,\cdots,\chi_{n+m-1}$ of the MR sKdV system. 
Assume the parities of $\mu_i$ are all odd, whereas among $\chi_i$, 
$\chi_0,\cdots,\chi_{n+[\frac{m+1}{2}]-1}$ are even and 
$\chi_{n+[\frac{m+1}{2}]},\cdots,\chi_{n+m-1}$ are odd. 
Then after performing the DB transformations triggered by these
(adjoint) wave functions, the transformed Lax operator reads
\bea
\hat{L}&=&Q_{(n,m)}LQ_{(n,m)}^{-1} \qquad 
Q_{(n,m)}=D^m+\sum_{i=0}^{m-1}d_iD^i+\sum_{i=0}^{n-1}e_iD^{-1}\mu_i
\label{qn}\\
&=&\pa^2+\hat{v}_1D+\hat{v}_0
\eea 
where the coefficients $d_i$ and $e_i$ are defined by
\be
Q_{(n,m)}\chi_i=0 
\qquad i=0,\cdots,n+m-1
\lb{linq}
\ee
and 
\bea
\hat{v}_1&=&(-1)^mv_1-2(d_{m-1})_x
\lb{qnv1}\\
\hat{v}_0&=&v_0-2(d_{m-2})_x-d_{m-1}((-1)^mv_1+\hat{v}_1)+
\frac{1-(-1)^m}{2}(Dv_1)
\lb{qnv0}
\eea
{\it Proof.\/} The gauge operator $Q_{(n,m)}$ can be realized easily
as follows: since  the number of wave functions is larger than that
of the adjoint wave functions thus part of 
even wave functions, say, $\chi_0,\cdots,\chi_{n-1}$
would be paired with odd adjoint wave functions $\mu_0,\cdots,\mu_{n-1}$
to form the binary DB transformation $R_n$ of Proposition 7. 
However, among the residual $m$ wave functions $\chi_n,\cdots,\chi_{n+m-1}$,
$[(m+1)/2]$ are even and $[m/2]$ are odd which will form the DB 
transformation $T_m$ of Proposition 3. Hence  
$Q_{(n,m)}=T_mR_n$ and its expression in (\ref{qn}) can be 
verified by induction. $\Box$

Notice that the parity of the gauge operator $Q_{(n,m)}$ depends on $m$, therefore
the case for $m$ being even or odd should be discussed separately.

For $m=2k$, it is convenient to define the following row vectors
\bea
\bchi^{[0]}&=&(\chi_0,\chi_1,\cdots,\chi_{n+k-1}) \qquad
\bchi^{[1]}=(\chi_{n+k},\cdots,\chi_{n+2k-1}) 
\lb{not1}\\
{\bf p\/}^{[0]}&=&(e_0, e_1,\cdots, e_{n-1},d_0,d_2,\cdots,d_{2k-2})\qquad
{\bf p\/}^{[1]}=(d_1,d_3,\cdots,d_{2k-1})\\
{\bf q\/}^{[i]}&=&\pax^k\bchi^{[i]}\qquad i=0,1
\eea
and matrices
\bea
\Om^{[0]}_{ij}&=&\Om(\mu_i,\chi_j)
\qquad i=0,\cdots,n-1; j=0,\cdots n+k-1\\
\Om^{[1]}_{ij}&=&\Om(\mu_i,\chi_{n+k+j})
\qquad i=0,\cdots,n-1; j=0,\cdots k-1\\
\Xi^{[i]}&=&
\left(
\ba{c}
\bchi^{[i]}\\
\pax\bchi^{[i]}\\
\vdots\\
\pax^{k-1}\bchi^{[i]}
\ea
\right)\qquad i=0,1
\lb{not9}
\eea 
where the superscript in bracket stands for the parity of the corresponding 
vector or matrix.

{\bf Proposition 9 :\/} After solving the linear equation (\ref{linq}) for $m=2k$, 
the transformed coefficients, $\hat{v}_1$ and $\hat{v}_0$, 
in Eqs.(\ref{qnv1})-(\ref{qnv0}) can be expressed as
\bea
\hat{v}_1&=&v_1+2(\frac{\det(\hat{D}-\hat{C}A^{-1}B)}{\det(D-CA^{-1}B)})_x
\label{ev1}\\
\hat{v}_0&=&v_0+2(\frac{\sdet \hat{{\bf M\/}}}{\sdet {\bf M\/}})_x-
(v_1+\hat{v}_1)(\frac{\det(\hat{D}-\hat{C}A^{-1}B)}{\det(D-CA^{-1}B)})
\label{ev0}
\eea
where the supermatrix 
${\bf M\/}=\left(
\ba{cc}
A & B\\
C & D
\ea
\right)
$
with 
\bea
A&=&
\left(
\ba{c}
\Om^{[0]} \\
\Xi^{[0]} 
\ea
\right)
\qquad
B=
\left(
\ba{c}
\Om^{[1]}\\
\Xi^{[1]}
 \ea
\right) \no\\
C&=&
(D\Xi^{[0]})
\qquad
D=(D\Xi^{[1]})
\eea
and its superdeterminant is defined by \cite{D}
\be
\sdet{\bf M\/}=\frac{\det(A-BD^{-1}C)}{\det D}=\frac{\det A}{\det(D-CA^{-1}B)}.
\ee
The matrix $\hat{\bf M\/}$ is constructed from ${\bf M\/}$ with its 
$(n+k)$-th row replaced by its $\pax$ derivation, whereas $\hat{D}$ and 
$\hat{C}$ are constructed from $D$ and $C$ with their last rows replaced 
by their $D$ derivation. 

{\it Proof.\/} For $m=2k$, the transformed coefficients $\hat{v}_1$ and $\hat{v}_0$
are expressed as
\bea
\hat{v}_1&=&v_1-2(d_{2k-1})_x\\
\hat{v}_0&=&v_0-2(d_{2k-2})_x+(v_1+\hat{v}_1)d_{2k-1}
\eea
Since the parity of the operator $Q_{(n,2k)}$ is even, 
we can rewrite the linear equation (\ref{linq}) as follows
\be
({\bf p\/}^{[0]},{\bf p\/}^{[1]}){\bf M\/}=-({\bf q\/}^{[0]},{\bf q\/}^{[1]})
\lb{linq2}
\ee
Now multiplying ${\bf M\/}^{-1}$ to the right of (\ref{linq2}), we have
\be
({\bf p\/}^{[0]},{\bf p\/}^{[1]})=-({\bf q\/}^{[0]},{\bf q\/}^{[1]})
\left(
\ba{cc}
(A-BD^{-1}C)^{-1} & -AB^{-1}(D-CA^{-1}B)\\
-D^{-1}C(A-BD^{-1}C)^{-1} & (D-CA^{-1}B)^{-1}
\ea
\right)
\ee
which implies 
\bea
{\bf p\/}^{[0]}(A-BD^{-1}C)
&=&-({\bf q\/}^{[0]}-{\bf q\/}^{[1]}D^{-1}C)\\
{\bf p\/}^{[1]}(D-CA^{-1}B)
&=&-({\bf q\/}^{[1]}-{\bf q\/}^{[0]}A^{-1}B)
\eea
Using the Cramer's rule, we obtain
\bea
d_{2k-2}&=&{\bf P\/}^{[0]}_{n+k}=-\frac{\sdet \hat{{\bf M\/}}}{\sdet {\bf M\/}} \\
d_{2k-1}&=&{\bf P\/}^{[1]}_k=
-\frac{\det(\hat{D}-\hat{C}A^{-1}B)}{\det(D-CA^{-1}B)}
\eea
which lead to the result (\ref{ev1})-(\ref{ev0}).$\Box$

For $m=2k+1$, some notations in Eqs.(\ref{not1})-(\ref{not9}) should be modified to
\bea
\bchi^{[0]}&=&(\chi_0,\chi_1,\cdots,\chi_{n+k})
\qquad \bchi^{[1]}=(\chi_{n+k+1},\cdots,\chi_{n+2k})
\lb{nnot1}\\
{\bf p\/}^{[0]}&=&(d_1,d_3,\cdots,d_{2k-1})\qquad
{\bf p\/}^{[1]}=(e_0,e_1,\cdots,e_{n-1},d_0,d_2,\cdots,d_{2k})\\
{\bf q\/}^{[i]}&=&(D^{2k+1}\bchi^{[i]})\qquad i=0,1  \\
\Om^{[0]}_{ij}&=&\Om(\mu_i,\chi_{j})\qquad i=0,\cdots,n-1; j=0,\cdots n+k\\
\Om^{[1]}_{ij}&=&\Om(\mu_i,\chi_{n+k+j+1})
\qquad i=0,\cdots,n-1; j=0,\cdots k-1\\
\Xi^{[i]}&=&
\left(
\ba{c}
\bchi^{[0]}\\
\pax\bchi^{[0]}\\
\vdots\\
\pax^k\bchi^{[0]}
\ea
\right)\qquad i=0,1
\lb{nnot9}
\eea
Then we can define
another supermatrix  
${\bf N\/}=\left(
\ba{cc}
\br{A} & \br{B}\\
\br{C} & D
\ea
\right)
$
where the matrices $\br{A}$ and  $\br{B}$ are constructed from the matrices
$A$ and  $B$ by taking into account the replacements (\ref{nnot1})-(\ref{nnot9}),
whereas $\br{C}$ does the same thing in addition to removing the last  row.

{\bf Proposition 10 :\/} Solving the linear equation (\ref{linq}) for $m=2k+1$, 
the transformed coefficients, $\hat{v}_1$ and $\hat{v}_0$, in 
Eqs.(\ref{qnv1})-(\ref{qnv0}) can be expressed as
\bea
\hat{v}_1&=&-v_1+2(\frac{\det(\ti{A}-\ti{B}D^{-1}\br{C})}
{\det(\br{A}-\br{B}D^{-1}\br{C})})_x
\lb{ov1}\\
\hat{v}_0&=&v_0+(Dv_1)+
2(\frac{\sdet {\bf N\/}}{\sdet \ti{\bf N\/}})_x+
(v_1-\hat{v}_1)(\frac{\det(\ti{A}-\ti{B}D^{-1}\br{C})}
{\det(\br{A}-\br{B}D^{-1}\br{C})})
\lb{ov0}
\eea
where $\ti{\bf N\/}$ is constructed from ${\bf N\/}$ with its last row replaced by
its $\pax$ derivation, whereas $\ti{A}$ and $\ti{B}$ are constructed from
$\br{A}$ and $\br{B}$ with their last rows replaced by 
their $D$ derivation. 

{\it Proof.\/} For $m=2k+1$, the linear equation (\ref{linq}) now becomes
\be
({\bf p\/}^{[1]},{\bf p\/}^{[0]}){\bf N\/}=-({\bf q\/}^{[1]},{\bf q\/}^{[0]})
\ee
Following the procedures for the case of $m=2k$, it is easy to show that
\bea
d_{2k}&=&{\bf p\/}^{[1]}_{n+k+1}=
\frac{\det(\ti{A}-\ti{B}D^{-1}\br{C})}{\det(\br{A}-\br{B}D^{-1}\br{C})}\\
d_{2k-1}&=&{\bf p\/}^{[0]}_k=-\frac{\sdet {\bf N\/}}{\sdet \ti{\bf N\/}}
\eea
which imply that the transformed coefficients $\hat{v}_1$ and $\hat{v}_0$
are given by (\ref{ov1}) and (\ref{ov0}), respectively.$\Box$

{\bf Remarks.\/} (1) In both cases ($m=2k$ or $2k+1$), 
the expressions for $\hat{v}_1$ and $\hat{v}_0$
are , indeed, independent  of the order of the even wave functions as well as the odd
ones. (2) For the case that the number of the adjoint wave functions is larger than 
that of the wave functions in a DBT, we can get a similar result just by exchanging 
the roles played by them.
(3)When we set $n=0$, the binary type expressions for $\hat{v}_1$ and $\hat{v}_0$
then reduce to the Wronskian type ones obtained in Ref.\cite{LM1}. Especially,
in the $m=2k$ case, it is possible to write $d_{2k-1}$ in terms of superdeterminant.

\section{Concluding remarks}
We have studied the solutions of the MR sKdV hierarchy by using DBTs. 
In addition to the previously known DBT \cite{L,LM1}, 
we provide the adjoint DBT which can be combined with the former one 
to form the binary DBT. Using these elementary DBTs, we obtain not only 
Wronskian type but also binary type solutions for the MR sKdV hierarchy. 
The super soliton solutions then can be constructed from the trivial
one which corresponds to $v_1=v_0=0$ by using the formulae 
derived in Sec. IV. Finally, we would like to remark that our approach involves 
only the algebra of S$\Psi$DOs, hence the formulation  is general enough to 
extend to the other cases such as super KP hierarchies and their reductions. 
We will leave these discussions to another 
publication.

{\bf Acknowledgments\/}
We would like to thank Q. P. Liu and M. Manas to point out an
error of the manuscript and inform us that the results of the 
propositions 6 and 7 were obtained in Ref.\cite{LM2} by using
the vectorial Darboux transformation.
This work is supported by the National Science Council of the
Republic of China under grant No. NSC-87-2811-M-007-0025.

\end{document}